\documentclass[pmlr]{jmlr}

\usepackage{longtable}

\usepackage{booktabs}
\usepackage[load-configurations=version-1]{siunitx} 

\theorembodyfont{\upshape}
\theoremheaderfont{\scshape}
\theorempostheader{:}
\theoremsep{\newline}

\jmlrvolume{...}
\jmlryear{...}
\jmlrworkshop{HEAR: Holistic Evaluation of Audio Representations}

\title[HEAR: Holistic Evaluation of Audio Representations]{HEAR: Holistic Evaluation of Audio Representations}

 \author{
 \Name{Joseph Turian} \Email{turian@gmail.com}\\
 \Name{Jordie Shier} \Email{jshier@uvic.ca}\\
 \Name{Humair Raj Khan} \Email{khumairraj@gmail.com}\\
 \AND
 \Name{Bhiksha Raj} \Email{bhiksha@cs.cmu.edu}\\
 \Name{Bj\"{o}rn W. Schuller} \Email{bjoern.schuller@imperial.ac.uk}\\
 \Name{Christian J. Steinmetz} \Email{c.j.steinmetz@qmul.ac.uk}\\
 \Name{Colin Malloy} \Email{malloyc@uvic.ca}\\
 \Name{George Tzanetakis} \Email{gtzan@ieee.org}\\
 \Name{Gissel Velarde} \Email{gv@urubo.org}\\
 \Name{Kirk McNally} \Email{kmcnally@uvic.ca}\\
 \Name{Max Henry} \Email{max.henry@mail.mcgill.ca}\\
 \Name{Nicolas Pinto} \Email{nicolas.pinto@gmail.com}\\
 \Name{Camille Noufi} \Email{cnoufi@stanford.edu}\\
 \Name{Christian Clough} \Email{christian.clough@gmail.com}\\
 \Name{Dorien Herremans} \Email{dorien.herremans@gmail.com}\\
 \Name{Eduardo Fonseca} \Email{eduardo.fonseca@upf.edu}\\
 \Name{Jesse Engel} \Email{jesseengel@google.com}\\
 \Name{Justin Salamon} \Email{salamon@adobe.com}\\
 \Name{Philippe Esling} \Email{philippe.esling@ircam.fr}\\
 \Name{Pranay Manocha} \Email{pmanocha@princeton.edu}\\
 \Name{Shinji Watanabe} \Email{swatanab@andrew.cmu.edu}\\
 \Name{Zeyu Jin} \Email{zejin@adobe.com}\\
 \Name{Yonatan Bisk} \Email{ybisk@cs.cmu.edu}\\
 }

\editor{Editor's name}

\begin{document}

\maketitle

\begin{abstract}
What audio embedding approach generalizes best to a wide range of downstream tasks across a variety of everyday domains without fine-tuning? The aim of the HEAR benchmark is to develop a general-purpose audio representation that provides a strong basis for learning in a wide variety of tasks and scenarios. 
HEAR evaluates audio representations using a benchmark suite across a variety of domains, including speech, environmental sound, and music. HEAR was launched as a NeurIPS 2021 shared challenge. In the spirit of shared exchange, each participant submitted an audio embedding model following a common API that is general-purpose, open-source, and freely available to use.
Twenty-nine models by thirteen external teams were evaluated on nineteen diverse downstream tasks derived from sixteen datasets. 
Open evaluation code, submitted models and datasets are key contributions, enabling comprehensive and reproducible evaluation, as well as previously impossible longitudinal studies. It still remains an open question whether one single general-purpose audio representation can perform as holistically as the human ear.

\end{abstract}

\section{Introduction}
\label{sec:intro}

The codification of strong general-purpose representations in natural language and computer vision has led to a renaissance in multimodal modeling and increased cross-discipline collaboration.  Audio is an equally rich source of information about the world, but outside of speech recognition it has not achieved the same degree of attention from the machine learning community.
This is a key challenge for the community, as good representations support good machine learning. And robust evaluation enables general representations.
Broad evaluation suites help prevent overfitting to common test sets \citep{Recht2018-qv} and have improved the state-of-the-art on language and vision representation learning \citep{Wang2018-xu,Wang2019-vw,Goyal2019-gw,Zhai2019-vl,DeYoung2019-fx}. 
In general practice, audio representations are not evaluated on a broad range of audio problems, 
and as a result, it is difficult to know which audio representation to use for a novel audio learning task.

The Holistic Evaluation of Audio Representations (HEAR) 
benchmark
was created to encourage the development of flexible audio representations, to give greater insight into how audio representations will generalize, and to enable fast development cycles both for researchers developing new models and researchers applying existing models. HEAR was launched as a NeurIPS 2021 shared challenge, and 
participants submitted audio representation models that are general-purpose, open-source, and freely available to use off-the-shelf. All HEAR compatible models follow a common API, which makes switching between models as simple as changing one line of code.

The HEAR benchmark includes nineteen tasks. During NeurIPS 2021, five were open tasks derived from three datasets for which the problem definition and evaluation data were available to participants, and 14 additional were secret tasks for evaluation, to which participants were completely blind.
While most of the tasks (open or secret) have good or promising solutions when worked on in isolation, the novelty of the HEAR benchmark is that the \emph{same} representation must be used to solve all of them. 
These tasks encompass multiple audio domains: speech, environmental sound, and music, with tasks that involve short and long time spans. 
HEAR datasets are easy to use: all are preprocessed to a common format with standard splits and self-explanatory human-readable metadata, and are distributed as tarfiles online.\footnote{\url{https://zenodo.org/record/5885750}} This alleviates the engineering effort required to work with datasets that require YouTube scraping, have variably documented preprocessing requirements, or are gatekept through closed-access request forms. Researchers are also welcome to use HEAR datasets under entirely open licenses (many of which allow commercial use), without using our downstream evaluation code.

Evaluation consists of classification tasks, both multiclass and multilabel, requiring either prediction over the entire audio scene (clip), or temporal-based onset detection of sound event \citep{sedeval}.
HEAR-compatible models can generate an embedding of arbitrary size, which is fed into a simple generic predictor by our open-source evaluation algorithm. 
%
%
Evaluation code, submitted models, and datasets are all available at \url{https://neuralaudio.ai/hear.html}.

\section{Background on representation learning}

At a high level, a learned representation (embedding) consists of a machine learning model 
that takes a low-level representation of the input and outputs a numerical representation, typically a fixed-size vector, that lends itself well to discriminative tasks (e.g., by training a simple MLP on these embeddings).
A good representation should (1) transfer to a wide range of different tasks and (2) transfer with limited supervision \citep{Goyal2019-gw,goyal2022vision}. 

In the following paragraphs, we describe trends from the natural language processing (NLP) and vision literature on representation learning, some of which have been applied to audio. Vision work is particularly relevant \citep{Amiriparian17-SSC}, as 2-D transformations of audio, such as the widely used log-Mel spectrogram \citep{davis1980comparison}, 
lend themselves well to methods designed to process 2-D input data.
For this reason, a common thread in the literature on audio representations is that vision models are applied to 2-D audio representations.
With that said, many of the insights from text-based language modeling, such as autoregressive neural modeling \citep{Bengio2000-vq},
predicting tokens masking as an unsupervised pretext task \citep{Collobert2011-qo},
and bidirectional transformers \citep{Devlin2018-uw},
have found their way into the audio literature, e.g., WaveNet \citep{Van_den_Oord2016-qr}, wav2vec \citep{Schneider2019-gx}, and HuBERT \citep{Hsu2021-jd}, respectively. Textless NLP like Generative Spoken Language Modeling (GSLM, \citet{Lakhotia2021-lo}), applies an NLP lens to spoken audio instead of written text.

\paragraph{Inducing representations}

The shallowest representation for audio is the raw digital audio signal itself. However, its extremely high dimensionality means it is rarely useful for discriminative tasks without additional processing, whether via manually crafted DSP engineering or transformations learned by training a neural network \citep{Trigeorgis16-AFE}. Better representations
might be obtained by applying a hand-crafted transformation based upon domain-expertise, such as the log-scaled Mel spectrogram \citep{davis1980comparison}, Mel Frequency Cepstral Coefficients (MFCC, \citet{Logan:MFCC:ISMIR:00}), constant Q-transform \citep{schorkhuber2010constant}, or scattering transform \citep{Anden2013-lb}. Audio filterbanks can also be learned \citep{Zeghidour2021-sk}.
Deep ML architectures can extract even more abstract, high-level representations \citep{aytar2016soundnet,Hershey2016-mc,Cramer2019-re}. Purely randomly weighted architectures impose particular inductive biases on data and can do better than hand-crafted baselines \citep{Saxe2011-uw,Pons2019-qq}.
However, it is more common to train these architectures.

\paragraph{Architectures}

The architecture of the model typically includes an encoder to transform the input, and can optionally also include temporal modelling to capture context, and/or a generative decoder. A common encoder architecture uses Convolutional Neural Networks (CNN) applied to a 2-D input \citep{Hershey2016-mc,Cramer2019-re}, or directly to the 1-D audio signal \citep{Van_den_Oord2016-qr,Baevski2020-sy}. Temporal context modelling is often achieved via Recurrent Neural Networks (RNN) \citep{Mehri2017-sr,Kalchbrenner2018-mi}, or Transformers \citep{Baevski2020-sy}. The latter, in particular, have achieved strong results for audio classification \citep{Gong2021-ast}, though they are costly to train from scratch. \citet{Koutini2021-gl} (\S\ref{sec:models}) demonstrate a faster training approach for audio transformer, which requires two GPU-days to pretrain on AudioSet. In reaction to the use of transformers, all-MLP architectures have demonstrated competitive results on language and vision tasks \citep{Liu2021-yg,Tolstikhin2021-pa}.

\paragraph{Training regimes}
Models can be trained on a (large-scale) supervised task, such as ImageNet \citep{imagenet} for vision and AudioSet \citep{Gemmeke2017-qe} for audio.
Multitask supervised training can further improve generalization \citep{Pascual2019-dr}.

To avoid the need for human-labeling, self-supervised models (a form of unsupervised learning) learn from large-scale unlabeled corpora. 
Many self-supervised approaches learn to correspond the original input with a different view on that same input, such as a semantically identical augmentation \citep{Chen2020-as,Tian2020-wm}.
To avoid collapsed solutions, self-supervised approaches historically used negative samples with a triplet loss \citep{chopra2005learning}, possibly requiring large negative batches \citep{Chen2020-as,saeed2021contrastive}, which can be expensive to train. Alternatives include quantization approaches to define uniform clusterings of representations \citep{Baevski2020-sy}, or carefully implemented asymmetric training architectures like BYOL \citep{Grill2020-wq,Niizumi2021-th} and SimSiam \citep{Chen2021-es}.
More recent are self-supervised approaches that avoid these aforementioned techniques, relying instead upon explicit and fundamental priors \citep{Zbontar2021-vk,Bardes2021-cl}.
Input augmentations can be used to increase the size of training data or provide corresponding views on the input \citep{Salamon:CNNAugmentationEnv:SPL:17}. \citet{fonseca2021unsupervised} and \citet{Wang2021-gk} discuss augmentations, including audio mixing, which \citet{Gong2021-kr,Wang2021-gw} explore in greater depth and argue is useful both for supervised and unsupervised regimes.

Multi-modal approaches learn the correspondence between different modalities of the input. Different modalities can accelerate compact learning in a single target modality by exploiting cross-modality structure. 
OpenL3 (\S\ref{sec:models}, \citet{Cramer2019-re}) is a broad-domain audio model trained on the correspondence between audio and video. Contrastive Language-Image Pre-training (CLIP, \citet{Radford2021-gt}) learns a model from 400\,M image-text pairs, and was successully applied on zero-shot tasks.
\citet{Wang2021-gk} contrastively induce audio representations from waveforms (1-D audio) and spectrograms, and \citet{Wang2021-gw} extend that to include correspondence with video frames.

Because pretraining large-scale models requires large quantities of data and can be computationally expensive, another research direction has been on distilling information from existing models that were trained on another modality for which more data are available. \citet{aytar2016soundnet} train SoundNET which distills audio representations from a pre-trained image classification model trained on large image datasets such as ImageNet \citep{imagenet}.
\citet{Ho-Hsiang_Wu_Prem_Seetharaman_Kundan_Kumar_Juan_Pablo_Bello_undated-ov}  (\S\ref{sec:models}) distill an audio representation (Wav2CLIP) from a large text-image model (CLIP) using video data to link the visual and audio modalities. 

\paragraph{Using and evaluating representations}

Representation models can be used in downstream tasks with full fine-tuning;
but the na\"{i}ve approach is simply to treat intermediate pre-trained model outputs as frozen embeddings, and this nonetheless provides a stark improvement over using raw features \citep{Turian2010-by}. 
%
%
Broad-scale evaluation of learned representations has been done in other ML domains, in NLP, for example: GLUE \citep{Wang2018-xu}, the harder SuperGLUE \citep{Wang2019-vw}, and ERASER \citep{DeYoung2019-fx}. Vision includes the FAIR self-supervision benchmark \citep{Goyal2019-gw} and VTAB \citep{Zhai2019-vl}.

\section{HEAR: Holistic Evaluation of Audio Representations}

A strong general-purpose audio representation should be as holistic as the human ear.
The goal of the HEAR competition is to evaluate audio representations across a variety of everyday domains, audio phenomena, with tasks that involve short and long time spans, sometimes with few labeled instances. Formal rules are provided on the HEAR website.\footnote{\url{https://neuralaudio.ai/hear2021-rules.html}}

\subsection{Related audio shared tasks}

Historical audio shared tasks, such as those from MIREX \citep{downie2014ten}, DCASE \citep{mesaros2017detection}, and INTERSPEECH ComParE \citep{Schuller13-TI2b}
have improved the community's understanding of audio modeling substantially. However, the bespoke nature of these tasks is a double-edged sword, requiring substantial custom tooling both by the challenge organizers and participants.
More recent audio shared tasks focus on reusability and generic task APIs.
SUPERB \citep{Yang2021-bk} focuses on a broad spectrum of speech tasks, and includes downstream evaluation ranging from simple classification to LSTM-based sequence modeling.
Although the Speech Commands v2 task is shared with HEAR, the other downstream tasks in SUPERB mainly deal with speech processing applications, including speech recognition, speaker verification, keyword spotting, etc., and these two evaluation activities are complementary to each other.
The NOn-Semantic Speech Benchmark (NOSS, \citet{Shor2020-kb}) comprises 6 paralinguistic tasks. Two tasks are shared with HEAR: CREMA-D and Speech Commands v2. Unfortunately, SAVEE and DementiaBank require filling out a request form, and VoxCeleb requires scraping YouTube.
HARES (Holistic Audio Representation Evaluation Suite)---not to be confused with our HEAR benchmark---is concurrently published work \citep{Wang2021-gy}.
HARES comprises 12 well-known downstream tasks including---like HEAR---ESC-50, Speech Commands v2, and an NSynth Pitch task, benchmarked on 13 models.
Where HARES differs from HEAR includes: a) HARES tasks are well-known benchmarks, whereas HEAR is a mix of well-known and novel benchmarks, b) HARES includes no few-shot tasks, all tasks have $\ge$ 2K samples, c) HARES results currently include no external submissions, d) evaluation code and dataset links are not provided and e) two of the tasks (AudioSet and VoxCeleb) tasks involve scraping YouTube. Datasets based upon YouTube require specialized code and lack reproducibility because videos are removed unpredictably \citep{Cramer2019-re}.
%
%
These generic audio evaluation suites, including our HEAR benchmark, intend to make it easy to evaluate existing models on novel tasks, at the expense of possible SOTA performance.

\subsection{Evaluation methodology}

Wrapping existing models into the HEAR API requires roughly 75 lines of code, much of which is boilerplate. New HEAR tasks can be run with no code changes.
HEAR includes two types of tasks:
1) Scene-based: Multi-class or multi-label classification of an entire audio clip;
2) Timestamp-based: Sound event detection/transcription, which involves detecting when exactly sound events occur over time by providing a start time, end time, and label for each sound event.
In both cases, the audio representation is frozen and used as the input feature vector to a shallow downstream MLP classifier, with no fine-tuning. Fine-tuning improves downstream performance (\citet{Baevski2020-sy,Shor2020-kb}), but increases training time. Crucially, the use of frozen embedings means that HEAR downstream evaluation code can be maintained solely in PyTorch, regardless of whether the embedding model was in TensorFlow or PyTorch.\footnote{We initially believed that imposing a restriction that all submitted models must be TensorFlow 2.x or PyTorch and pip3-installable would facilitate easy orchestration of model testing. However, models submitted with competing TensorFlow, CUDA, CuDNN, and pypi dependencies lead us to suggest that future ML challenge organizers standardize on the latest stable microversion of all deep learning packages.}

A timestamp-based task can be simplified to a frame-based sequence-labeling task of the audio at regular intervals \citep{Kelz2016-vy}, and we use a common postprocessing step to compose predictions from multiple timesteps and extract discrete labeled events with start and ends times \citep{sedeval}.
Framewise accuracy (the decomposed multilabel prediction, computed at regular timesteps) does not always correlate well with the perceptual quality of event-onset FMS \citep{Hawthorne2017-ag} because they ignore the interplay between the frame representations and more sophisticated downstream inference \citep{Cheuk2021-rl}.
See \sectionref{apd:downstream} for details on the downstream training regime.


\subsection{Evaluation tasks}

The following are the HEAR evaluation tasks.
For simplicity and reproducibility, we have preprocessed each relevant datasets to all commonly used sample rates (16000, 22050, 32000, 44100), fixed the length of the audio clips, predefined training splits, and packaged each dataset in a self-explanatory common format with human-readable metadata.
They all have open licenses (some of which permit commercial use), with the exception of the GTZAN corpora which are widely used but of unknown license status.
We encourage the community to benchmark on HEAR datasets, even if they do not follow the HEAR rules or HEAR API.


Open tasks were released early in the NeurIPS 2021 shared challenge, to encourage participation and to allow participants to debug and refine their submissions: Speech Commands v2 (full and 5h versions), NSynth Pitch (50h and 5h versions), and DCASE 2016 Task 2.
Tasks are summarized in \tableref{tab:datasets} described with more detail in \tableref{tab:task-summary} and \sectionref{apd:datasets}.

\begin{table}[htbp]
\floatconts
  {tab:datasets}%
  {\caption{HEAR tasks.}}%
  {
\raggedright
{\bfseries Speech Commands (version 2), 5h and full}
Spoken commands classification.
\\
{\bfseries NSynth Pitch, 5h and 50h}
Pitch classification of synthesized sounds.
\\
{\bfseries DCASE 2016 Task 2}
Office sound event detection in synthesized scenes.
\\
{\bfseries Beehive States}
Binary classification of normal vs.\ queen-less beehives.
\\
{\bfseries Beijing Opera Percussion}
Classification of six Beijing Opera percussion instruments.
\\
{\bfseries CREMA-D}
Speech emotion recognition.
\\
{\bfseries ESC-50}
Environmental sound classification.
\\
{\bfseries FSD50K}
Broad-domain audio multi-labeling.
\\
{\bfseries Gunshot Triangulation}
Identify location of microphone recording a gunshot, using classification.
\\
{\bfseries GTZAN Genre}
Music genre classification.
\\
{\bfseries GTZAN Music Speech}
Classification of audio into music or speech.
\\
{\bfseries LibriCount}
Multiclass speaker count identification.
\\
{\bfseries MAESTRO 5h}
Music transcription.
\\
{\bfseries Mridingham Stroke and Mridingham Tonic}
Non-Western pitched percussion. Classification of stroke or tonic.
\\
{\bfseries Vocal Imitations}
Match a vocal imitation to the type of sound imitated, using classification.
\\
{\bfseries VoxLingua107 Top 10}
Spoken language identification.
  }
\end{table}

\section{Models evaluated}\label{sec:models}

Evaluated models are described below. \tableref{tab:model-summary} summarizes model properties.
HEAR began with three strong baseline models (\S\ref{sec:models:baseline}), each pretrained on a
different audio domain.
We report on 13 external teams' submissions to the HEAR NeurIPS 2021 shared challenge (\S\ref{sec:models:submitted}). 



\subsection{Baseline models}\label{sec:models:baseline}

\paragraph{wav2vec2} wav2vec2 (1-D CNN and positional transformer) \citep{Baevski2020-sy}. Self-supervised pretraining on 100K hours
of speech from VoxPopuli \citep{Wang2021-qv}.

\paragraph{CREPE}
1-D CNN. Supervised pretraining of pitch-tracking on 16 hours of synthesized music. \citep{Kim2018-gt}

\paragraph{OpenL3}
2-D CNN. Multi-modal contrastive self-supervised pretraining of audio/video correspondence on 6K hours of AudioSet broad-domain YouTube content. (\citet{Cramer2019-re}, earlier \citet{arandjelovic2017look})
HEAR implementation by Jon Nordby.

\subsection{Submitted models}\label{sec:models:submitted}

\paragraph{AMAAI Lab SUTD wav2vec2+DDSP}

An ensemble of wav2vec2~\citep{Baevski2020-sy} and two DDSP encoders \citep{engel2020ddsp}. The wav2vec2 model is pretrained on the  Librispeech~\citep{librispeech} and MAESTRO~\citep{Hawthorne2018-yb} datasets. One DDSP encoder is CREPE, the other is a non-pretrained loudness encoder.

\paragraph{AMAAI wav2vec2 music+speech}
wav2vec2 model~\citep{Baevski2020-sy}. Pretrained on Librispeech~\citep{librispeech} and MAESTRO~\citep{Hawthorne2018-yb}.

\paragraph{CP-JKU PaSST base, base2level, base2levelmel} 
Patchout fast (2-D) spectrogram transformer (PaSST, \citet{Koutini2021-gl}).
Initialized from a ImageNet vision transformer model, and further pretrained on 10s audio from AudioSet to perform supervised tagging. base2level concatenates a longer window (160 ms and 800ms) for timestamp embeddings. base2levelmel 
additionally concatenates the raw melspectrogram as well.

\paragraph{CVSSP (University of Surrey) PANNs}
2-D CNN14.
Pretrained on AudioSet with supervision \citep{IqbalTurab2020PLPA}.

\paragraph{Descript/MARL Wav2CLIP}
2-D ResNet18. Pretrained multimodally using contrastive learning on the 600h VGGSound corpus \citep{Chen2020-sm} (without supervised labels) to distill the Contrastive Language-Image Pre-training (CLIP, \citet{Radford2021-gt}) language and image model to a corresponding audio embedding 
\citep{Ho-Hsiang_Wu_Prem_Seetharaman_Kundan_Kumar_Juan_Pablo_Bello_undated-ov}.

\paragraph{IUT-CSE kwmlp and audiomlp}
Sequentially stacked gated MLP model \citep{Liu2021-yg}, taking (2-D) MFFCs as input. kwmlp \citep{morshed2022-hear1} is pretrained with supervision on Speech Commands v2. audiomlp is pretrained with supervision on HEAR open task datasets: Speech Commands v2, DCASE 2016 Task 2, and NSynth Pitch.

\paragraph{Kuroyanagi hearline}
2-D conformer model. Pretraining unknown.

\paragraph{Logitech AI SERAB BYOL-S}
2-D CNN. Self-supervised pretraining using the BYOL self-supervised approach \citep{Grill2020-wq} adapted to audio (BYOL-A, \citet{Niizumi2021-th}), pretrained on the speech subset of AudioSet \citep{elbanna2022-hear1}.

\paragraph{NTU-GURA (fusion) avg/cat hubert/wav2vec2/crepe}
Three models (HuBERT \citet{Hsu2021-jd}, wav2vec2, CREPE) combined in a variety of ways: averaged or concatenated \citep{Wu2022-hear1}. Fusion of multiple model layers was optionally included. fusion\_cat\_xwc\_time is a variation of fusion\_cat\_xwc with a different approach to matching timestamps when concatenating different models' emmbeddings.

\paragraph{RedRice/Xiaomi EfficientNet-B2}
2-D EfficientNet-B2 \citep{Tan2019-em}. Pretrained on supervised AudioSet tags.
Instead of global averaging pooling, decision-level pooling is used.
Timestamp embeddings are smeared scene embeddings.

\paragraph{Sony UDONS ViT}
Vision transformer (ViT, \citet{Kolesnikov2021-vit}). Pretrained on 360h of Librispeech to predict the correct permutation \citep{Noroozi2016-ul,Carr2021-fv} of up to 5 patches of mel-spectrogram input, shuffled in time.

\paragraph{Soundsensing YAMNet}

2-D MobileNet \citep{Howard2017-us}. Pretrained to tag AudioSet.

\paragraph{Stellenbosch LSL Audio DBERT}

1-D CNN encoder and modified BERT transformer. Pretrained as the discriminator with a GAN objective, using the clustering model as the generator, on 960 hours of Librispeech \citep{librispeech}.
Embeddings are taken from layer 16 of 24 by default.

\section{Results and Discussion}

In \figureref{fig:leaderboard} we present the primary score of submitted models on each HEAR task. By default, evaluation uses a deterministic seed, for reproducibility. Nonetheless, scores are stable across our evaluation, with a median 95\% confidence interval of 2.5e-3 when seeding of model weights and hyperparameter grid points is selected non-determinisically.
\citet{Shor2020-kb,Ho-Hsiang_Wu_Prem_Seetharaman_Kundan_Kumar_Juan_Pablo_Bello_undated-ov} present scores for some of the the same models and tasks. HEAR reported scores are similar but not identical, due to downstream training differences.

\begin{figure}[hbtp]
\floatconts
  {fig:tsnes}
  {\caption{t-SNE visualizations of tasks and models, based upon normalized scores. Missing normalized scores were imputed using sklearn's multivariate IterativeImputer.}}
  {%
    \subfigure[Tasks]{\label{fig:tasks-tsne}%
      \includegraphics[width=0.6\linewidth]{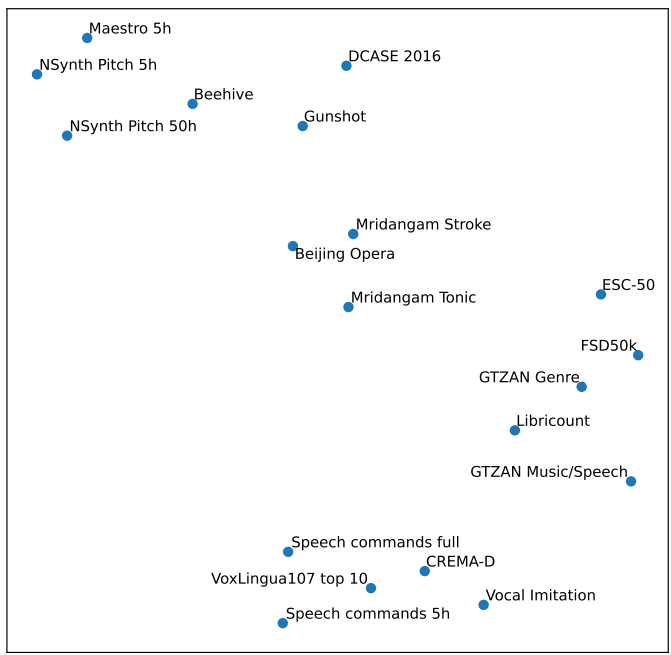}}%
    \par\bigskip
    \subfigure[Models]{\label{fig:models-tsne}%
      \includegraphics[width=0.6\linewidth]{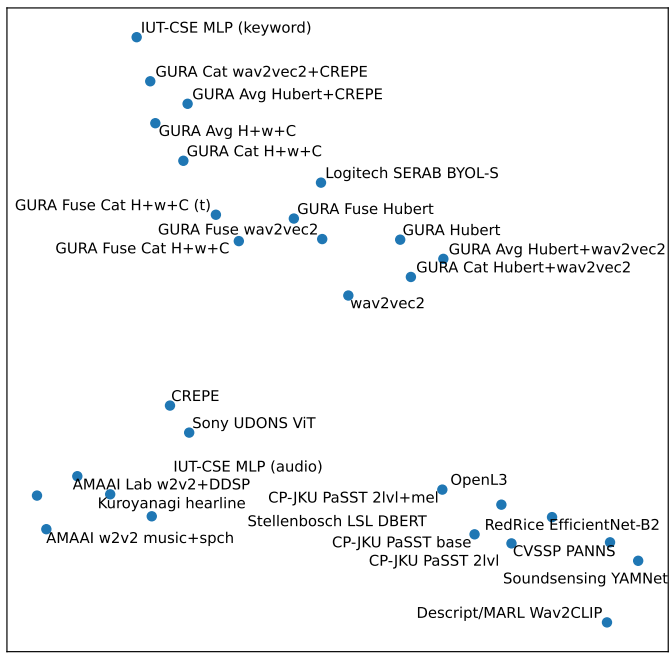}}%
  }
\end{figure}

To display model similarity at a glance, we present t-SNE visualizations of normalized scores by task (\figureref{fig:tasks-tsne}) and by model (\figureref{fig:models-tsne}).
We also show correlation tables for tasks (\figureref{fig:task-correlation}) and models (\figureref{fig:model-correlation}) to give greater insight into model and task similarity, in similar spirit to the confusion matrices of \citet{Ho-Hsiang_Wu_Prem_Seetharaman_Kundan_Kumar_Juan_Pablo_Bello_undated-ov}.
%
%
\citet{Zhai2019-vl} compare a variety of aggregation techniques for evaluating cross-task model performance, and find that they are all highly correlated, settling upon simple mean top-1. \citet{Gosiewska2020-hv} proposes an ELO-like meta-score for cross-task model performance, similar to a chess rating. Although it is tempting to give a single score for every model, we believe that would strip out important nuances shown in the full score table \citep{DeYoung2019-fx}.

For these summary figures, we normalize each model/task score. Normalized scores allow us to compare models and tasks against each other, under the assumption each task is equally weighted. The normalization procedure is as follows:
1) For each task, we standardize the scores to zero mean and unit variance. Unlike transforming tasks to ranks, we assume that the scale of intra-task scores is important.
2) The standardized scores are Winsorized (clamped) to have variance within [-1, +1]. By limiting the importance of extremely high or low scores on a single task, this approach allows for better inter-task comparison.


\begin{figure}[htbp]
\floatconts
    {fig:leaderboard}%
    {\caption{Primary score of submitted models on each HEAR task. Normalized scores are used to show the heat-value of each cell. Missing cells indicate that the model did not successfully complete the task (exhausting GPU memory or exceeding 24 hours downstream training time).
             }}%
    {\includegraphics[width=\textwidth]{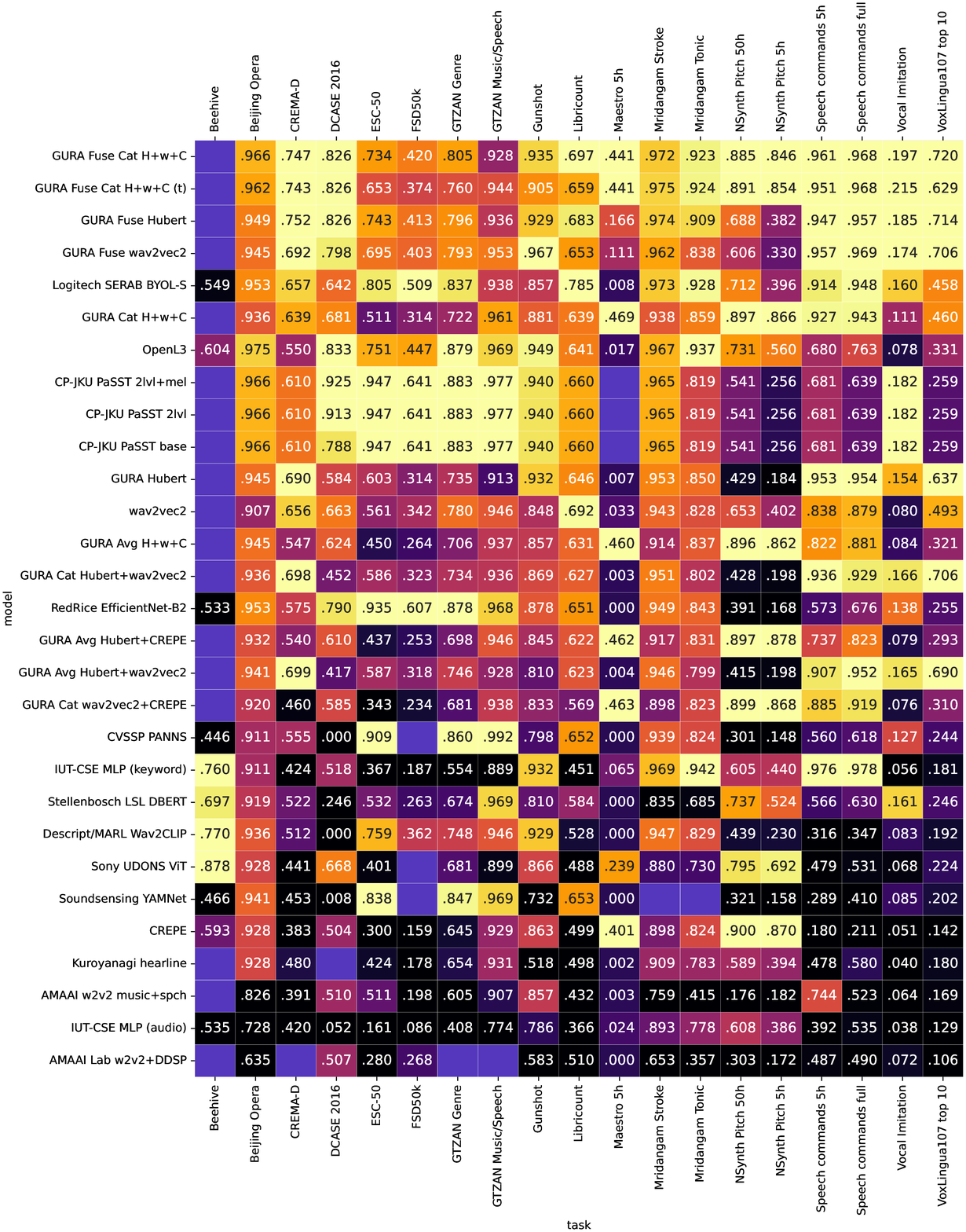}}
\end{figure}




In the following paragraphs, we describe a few interesting patterns and trends in the submitted models. Many evaluted models use the last layer as the representation. It is known that non-final layers and/or fusing various layers might capture more information
\citep{Shor2020-kb,alexeib2020-fusion,Yang2021-bk}. Intermediate layers often model audio phenomena that are not necessary for the final loss.
NTU-GURA's ablation studies support that, as evidence by the relative performance of their different models. For conciseness, we use the term ``strong speech models'' to refer to NTU-GURA's fused models that include pretrained speech models.

\paragraph{Pitch tasks} NSynth pitch and Maestro tasks have similar results, and models that include CREPE embeddings \citep{Kim2018-gt} perform best. This makes sense as these tasks require modeling of pitch, which CREPE was specifically trained for, while many other representations focus on discriminating between semantic objects (e.g., cat vs dog or guitar vs piano) but are pitch agnostic. 
Interestingly, models trained for semantic discrimination (e.g., via AudioSet) and speech models do nonetheless represent pitch to some degree, as evidenced by the decent performance of OpenL3 and wav2vec2 on these tasks.

\paragraph{Broad Domain Semantic-Object Tagging} FSD50K and ESC-50 semantic-object tagging results are strongly correlated, as well as---perhaps surprisingly---GTZAN genre tagging.
The models that perform the best on this group are the ones pretrained on the AudioSet semantic-object tagging task. 
What we glean from this large-scale survey of diverse models is that results on ESC-50 and GTZAN genre tagging are strongly predictive of results on the more nuanced FSD50K task, despite being an order of magnitude smaller and not using the corrected GTZAN artist-conditional splits from \citep{Sturm13-gtzan}, suggesting faster inroads for research iteration.
One valuable point-based contribution of HEAR is that the CP-JKU PaSST models achieve a new state-of-the-art on FSD50K despite no fine-tuning, a mean average precision (mAP) of 0.641 on FSD50K, compared to the recent literature \citep{Gong2021-kr,Ho-Hsiang_Wu_Prem_Seetharaman_Kundan_Kumar_Juan_Pablo_Bello_undated-ov,fonseca2021shift}. 


\paragraph{Vocals} FSD50K scores are also similar to those of Vocal Imitations and LibriCount. 
This is perhaps because Vocal Imitations comprises broad non-semantic vocalizations and LibriCount involves detecting multiple simultaneous audio events. The strong speech and PaSST models do the best on Vocal Imitations. On LibriCount, SERAB BYOL-S does the best as a non-semantic speech model, with decent performance from strong speech models. 

\paragraph{Speech} As we move into the speech domain, LibriCount and Vocal Imitations have the most similarity to CREMA-D emotion detection, which then is most similar to VoxLingua107 Top 10 language identification, which in turn is correlated with Speech Commands, following a trend from ``environmental'' to paralinguistic to semantic. The strong speech models do the best on these tasks.


What is most interesting about our diverse survey of 29 models $\times$ 19 tasks is, perhaps, the most difficult to explain results: tasks that defy neat categorization suggest the fragile, unpredictable boundaries of existing models. DCASE 2016 Task 2 seems {\em a priori} similar to FSD50K and ESC-50, but not in practice. Vocal Imitations are human-depictions of all kinds of sounds. Gunshot Triangulation is an extremely low-resource task with only 88 instances. Beijing Opera and Mridingham Stroke and Tonic are non-Western music tasks. For these tasks, our contribution is a negative result: we have no simple story or obvious pretraining data to attack them. Robust generalization of 
$>$10-billion-parameter models from NLP \citep{Brown2020-it} and vision \citep{goyal2022vision} suggest one path forward.

\section{Conclusion}

General-purpose models that transfer to few-shot and zero-shot scenarios are highly desirable. The audio community has followed the NLP and vision communities in using increasingly sophisticated representation learning approaches. The HEAR benchmark allows the audio community also to follow the trend of broad-scale reproducible evaluation.

HEAR is about openness. The datasets and the submissions are as open as possible.
All HEAR datasets are preprocessed to a common format with standard splits, and distributed as tarfiles. This alleviates the risk of dataset rot common in YouTube scraping, 
and the difficulty of acquiring data locked behind 
closed-access request forms.
All HEAR submissions have code that is Apache 2.0 compatible, models that are CC-Attribution compatible, and follow a common API, so switching between them requires a single line of code. Evaluation code, submitted models, and datasets are key contributions of HEAR, available at \url{https://neuralaudio.ai/hear.html}.

Twenty-nine models were evaluated on 19 diverse downstream tasks, spanning speech, environmental sounds, and music, and datasets that don't fit neatly into any rubric, as well as datasets that span the boundaries of multiple audio domains.
This large standardized set of tasks and models pave the way for comprehensive and reproducible evaluation, enabling previously impossible longitudinal studies.
We are eager to help onboard new tasks into the HEAR benchmark suite, particularly
unusual and/or few-shot audio tasks.
The largest-scale HEAR scene-embedding tasks and the CPU-gated evaluation of timestamp-embedding tasks were the most difficult tasks to run, sometimes requiring 24 hours for downstream eavluation of a single model-task pair on an A100 GPU, despite no fine-tuning.

Before an evaluation like HEAR, it would be easy for the community to suggest which audio tasks are predictably hard: large-scale, well-defined datasets with no more low-hanging fruit that are known to be difficult to hill-climb. Our contribution---the existence and easy accessibility of HEAR datasets, models, and evaluation code---allows the community to probe what we don't know. And the central question posed by HEAR remains open: Can one single general-purpose audio representation perform as holistically as the human ear? If one does, then there is clearly more work to be done towards achieving it.

\acks{HEAR was sponsored by Google, and competition evaluation was performed on Google Cloud Platform.}

\bibliography{hear2021}

\appendix

\newpage



\begin{table}[tbhp]
\floatconts
  {tab:task-summary}%
  {\caption{Summary of the 19 evaluation tasks of HEAR. Includes the embedding type (timestamp (T) or scene (S)), the predictor type (multiclass (C) or multilabel (L)), the split method used during downstream evaluation (train/validation/test (TVT) or $K$-fold), the duration of clips in seconds, the total number of clips for each task, the primary evaluation metric, and whether or not the task is novel. Novel tasks are not comparable to the literature. For all tasks except FSD50k, clips were standardized to one length using padding or trimming, typically the 95th percentile length in the original corpus.}}%

\scriptsize
    \begin{tabular}{lcccrrcc}
        \toprule
        \bfseries Task Name & \bfseries Embed & \bfseries Predictor & \bfseries Split & \bfseries Duration & \bfseries \# clips & \bfseries Evaluation & \bfseries Novel \\
       	& \bfseries Type & \bfseries Type & \bfseries Method & (seconds) & & \bfseries Metric & \\
        \midrule
        \bfseries Open Tasks & & & & & & &        \vspace{1mm} \\
        DCASE 2016 Task 2 & T & L & TVT & 120.0 & 72 & Onset FMS & \checkmark\\
        NSynth Pitch 5hr & S & C & TVT & 4.0 & 5000 & Pitch Acc. & \checkmark \\
        NSynth Pitch 50hr & S & C & TVT & 4.0 & 49060 & Pitch Acc. & \checkmark \\
        Speech Commands 5hr & S & C & TVT & 1.0 & 22890 & Accuracy & \checkmark \\
        Speech Commands Full & S & C & TVT & 1.0 & 100503 & Accuracy & \\
        \midrule
        \bfseries Secret Tasks & & & & & &         \vspace{1mm} \\
        Beehive States & S & C & TVT & 600.0 & 576 & AUCROC \\
       	Beijing Opera Percussion & S & C & 5-fold & 4.77 & 236 & Accuracy & \checkmark \\
       	CREMA-D & S & C & 5-fold & 5.0 & 7438 & Accuracy & \\
       	ESC-50 & S & C & 5-fold & 5.0 & 2000 & Accuracy & \\
       	FSD50K & S & L & TVT & 0.3 - 30.0 & 51185 & mAP & \\
       	Gunshot Triangulation & S & C & 7-fold & 1.5 & 88 & Accuracy & \checkmark \\
       	GTZAN Genre & S & C & 10-fold & 30.0 & 1000 & Accuracy & \\
       	GTZAN Music Speech & S & C & 10-fold & 30.0 & 128 & Accuracy & \\
       	LibriCount & S & C & 5-fold & 5.0 & 5720 & Accuracy & \\
       	MAESTRO 5hr & T & L & 5-fold & 120.0 & 185 & Onset FMS  & \checkmark \\
        Mridangam Stroke & S & C & 5-fold & 0.81 & 6977 & Accuracy & \checkmark \\
        Mridangam Tonic & S & C & 5-fold & 0.81 & 6977 & Accuracy & \checkmark \\
        Vocal Imitations & S & C & 3-fold & 11.26 & 5601 & mAP & \checkmark \\
     	VoxLingua107 Top10& S & C & 5-fold & 18.64 & 972 & Accuracy & \checkmark \\
        \bottomrule
    \end{tabular}
\end{table}

\section{Evaluation Tasks}\label{apd:datasets}

Our 19 tasks were derived from 16 datasets, as described in more detail below.
Tasks described as ``novel'' are not comparable to the literature. A summary of task statistics is available in \tableref{tab:task-summary}.

\paragraph{Speech Commands (version 2), 5h and full}
Classification of known spoken commands, with additional categories for silence and unknown commands \citep{Warden2018-he}.
As per the literature, models are evaluated by prediction accuracy. We also provide a 5-hour subset of the training data. We use the predefined train and test split, and note that the test data has a different distribution of labels from the training data.

\paragraph{NSynth Pitch, 5h and 50h}
NSynth Pitch is a novel multiclass classification problem. The goal of this task is to classify instrumental sounds from the NSynth Dataset \citep{Engel2017-sz} into one of 88 pitches. Results for this task are measured by pitch accuracy, as well as chroma accuracy. Chroma accuracy considers only the pitch ``class'' i.e., pitches that are a multiple-of-octaves apart are considered equivalent.
For HEAR we created two versions of this dataset: a 5 hour and 50 hour version.
Unlike \citet{Carr2021-fv}, we treat this as a classification, not regression problem. 

\paragraph{DCASE 2016 Task 2}
A novel office sound event detection in synthesized scenes, adapted from DCASE 2016 Task 2 \citep{Mesaros2018-ya}. Novel, insofar as our evaluation uses different splits. The original imbalanced splits did not work well our generic cross-validation.

Postprocessing: Predictions were postprocessed using 250 ms median filtering. At each validation step, a minimum event duration of 125 or 250\,ms was chosen to maximize onset-only event-based F-measure (with 200\,ms tolerance). Scores were computed using \texttt{sed\_eval} \citep{sedeval}.

\paragraph{Beehive States}
This is a binary classification task using audio recordings of two beehives \citep{Nolasco2018-gw}. The beehives are in one of two states: a normal state, and one in which the queen bee is missing (``queen-less''). At 10 minutes long, this task has the longest audio clips in HEAR.

\paragraph{Beijing Opera Percussion}
This is a novel audio classification task developed using the Beijing Opera Percussion Instrument Dataset \citep{Tian2014-nw}. The Beijing Opera uses six main percussion instruments that can be classified into four main categories: Bangu, Naobo, Daluo, and Xiaoluo.

\paragraph{CREMA-D}
CREMA-D is a dataset for emotion recognition \citep{Cao2014-hi}. The original dataset contains audiovisual data of actors reciting sentences with one of six different emotions (anger, disgust, fear, happy, neutral and sad). For HEAR, we only use the audio recordings (which differs from much but not all of the literature).

\paragraph{ESC-50}
This is a multiclass classification task on environmental sounds. The ESC-50 dataset is a collection of 2000 environmental sounds organized into 50 classes \citep{Piczak2015-bl}. Scores are averaged over 5 folds. (The folds are predefined in the original dataset.)

\paragraph{FSD50K}
FSD50K is a multilabel task \citep{Fonseca2020-xq}. This dataset contains over 100 hours of human-labeled sound events from Freesound (\url{https://freesound.org/}). Each of the $\approx$51\,k audio clips is labeled using one or more of 200 classes from the AudioSet Ontology, encompassing environmental sounds, speech, and music. Unlike the other datasets, for FSD50K scene embeddings we did not alter the audio clip length. Each clip is between 0.3 and 30 seconds long. We use the predefined train/val/eval split. Evaluation is done using mean average precision (mAP).

\paragraph{Gunshot Triangulation}
Gunshot triangulation is a novel resource multiclass classification task that utilizes a unique dataset: gunshots recorded in an open field using iPod Touch devices \citep{cooper_seth_2020_3997406}. This data consist of 22 shots from 7 different firearms, for a total of 88 audio clips, the smallest dataset in HEAR. Each shot is recorded using four different iPod Touches, located at different distances from the shooter. The goal of this task is to classify audio by the iPod Touch that recorded it, i.e., to identify the location of the microphone.
The dataset was split into 7 different folds, where each firearm belonged to only one fold. Results are averaged over each fold.

\paragraph{GTZAN Genre}
The GTZAN Genre Collection \citep{Tzanetakis2002-wm} is a dataset of 1000 audio tracks (each 30 seconds in duration) that are categorized into ten genres (100 tracks per genre). The task is multiclass classification. As per the literature, scores are averaged over 10 folds. However, we don't used the corrected artist-conditional splits from \citep{Sturm13-gtzan}.

\paragraph{GTZAN Music Speech}
\href{http://marsyas.info/downloads/datasets.html#music-speech}{GTZAN Music Speech} is a binary classification task, where the goal is to distinguish between music and speech. The dataset consists of 120 tracks (each 30 seconds in duration) and each class (music/speech) has 60 examples.


\paragraph{LibriCount}
LibriCount is a multiclass speaker count identification task \citep{Stoter2018-ec}. The dataset contains audio of a simulated cocktail party environment with between 0 to 10 speakers. The goal of this task is to classify how many speakers are present in each of the recordings. Following \citet{Stoter2018-xq}, we treat this as a classification, not regression, problem.

\paragraph{MAESTRO 5h}
This is a novel music transcription task adapted from MAESTRO. For HEAR, we created a subsampled version that includes 5 hours of training and validation audio, in 120 second clips. To evaluate submissions, a shallow transcription model was trained on timestamp-based embeddings provided by the participant models.

We use note onset FMS and note onset with offset FMS for evaluation, as per the original MAESTRO paper \citep{Hawthorne2018-yb} and the preceding Onsets and Frames paper \citep{Hawthorne2017-ag}.

Note onset measures the ability of the model to estimate note onsets with 50\,ms tolerance and ignores offsets. Note onset w/ offset includes onsets as well as requires note duration within 20\% of ground truth or within 50\,ms, whichever is greater.

\paragraph{Mridingham Stroke and Mridingham Tonic}
We used the Mridangam Stroke Dataset \citep{Anantapadmanabhan2013-wf} for two novel multiclass classification tasks: Stroke classification and Tonic classification. The Mridingam is a pitched percussion instrument used in carnatic music, which is a sub-genre of Indian classical music. This dataset comprises 10 different strokes played on Mridingams with 6 different tonics.

\paragraph{Vocal Imitations}
Vocal Imitations \citep{Kim2018-ry} is a novel multiclass classification task, where the goal is to match a vocal imitation of a sound with the sound that is being imitated. The dataset contains 5601 vocal imitations of 302 reference sounds, organized by AudioSet ontology. Given a vocal sound, the classification task is to retrieve the original audio it is imitating.

\paragraph{VoxLingua107 Top 10}
This is a novel multiclass classification task derived from the VoxLingua107 dataset \citep{Valk2021-ar}. The goal of the task is to identify the spoken language in an audio file. For HEAR we selected the top 10 most frequent languages from the development set, which resulted in just over 5 hours of audio over 972 audio clips.


\begin{table}[htbp]
\floatconts
  {tab:model-summary}%
  {\caption{Properties of baseline and submitted models, including: whether the model processes raw audio (1-D) or spectrograms (2D); on what kind of data the model is pretrained; the number of million parameters; the size of the output embedding for scene and timestamp tasks; and the number of minutes the model spends embedding Speech Commands V2. We caution that embedding time is not the entire picture, if participants did not do simple speed optimizations. For example, the CREPE wrapper (also used by GURA) is known not to exploit GPU batch parallelism.}}%
  {
\scriptsize
\begin{tabular}{lcccccrrrrr}
\toprule
\bfseries & \multicolumn{2}{|c|}{\bfseries Input} & \multicolumn{3}{|c|}{\bfseries Pretraining data} & \bfseries \# M & \multicolumn{2}{|c|}{\bfseries Embed dim} & \bfseries Time \\
\bfseries Model & \bfseries 1D & \bfseries 2D &  \rotatebox{0}{\bfseries \tiny speech} & \rotatebox{0}{\bfseries \tiny broad} & \rotatebox{0}{\bfseries \tiny music} & \bfseries params & \bfseries scene & \bfseries time & \bfseries min \\
\midrule
\footnotesize OpenL3 & & \checkmark & & \checkmark & & 4.7 & 512 & 512 & 94.9 \\
\footnotesize wav2vec2 & \checkmark & & \checkmark & & & 315.4 & 1024 & 1024 & 8.9 \\
\footnotesize CREPE & \checkmark & & & & \checkmark & 22.2 & 2048 & 2048 & 38.3 \\
\midrule
\footnotesize AMAAI Lab wav2vec2+DDSP & \checkmark & & \checkmark & & \checkmark & 98.8 & 871 & 871 & 43.6 \\
\footnotesize AMAAI wav2vec2 music+speech & \checkmark & & \checkmark & & \checkmark & 300.0 & 768 & 768 & 5.0 \\
\footnotesize CP-JKU PaSST 2lvl & & \checkmark & & \checkmark & & 86.2 & 1295 & 2590 & 14.5 \\
\footnotesize CP-JKU PaSST 2lvl+mel & & \checkmark & & \checkmark & & 86.2 & 1295 & 3358 & 5.8 \\
\footnotesize CP-JKU PaSST base & & \checkmark & & \checkmark & & 86.2 & 1295 & 1295 & 5.8 \\
\footnotesize CVSSP PANNS & & \checkmark & & \checkmark & & 80.8 & 2048 & 2048 & 3.9 \\
\footnotesize Descript/MARL Wav2CLIP & & \checkmark & & \checkmark & & 11.7 & 512 & 512 & 3.1 \\
\footnotesize GURA Avg H+w+C & \checkmark & & \checkmark & & \checkmark & 1339.0 & 1024 & 1024 & 40.0 \\
\footnotesize GURA Avg Hubert+Crepe & \checkmark & & \checkmark & & \checkmark & 1022.0 & 1024 & 1024 & 33.9 \\
\footnotesize GURA Avg Hubert+wav2vec2 & \checkmark & & \checkmark & & & 634.0 & 1024 & 1024 & 14.6 \\
\footnotesize GURA Cat H+w+C & \checkmark & & \checkmark & & \checkmark & 1339.0 & 3072 & 3072 & 40.1 \\
\footnotesize GURA Cat Hubert+wav2vec2 & \checkmark & & \checkmark & & & 634.0 & 2048 & 2048 & 14.4 \\
\footnotesize GURA Cat wav2vec2+crepe & \checkmark & & \checkmark & & \checkmark & 339.0 & 2048 & 2048 & 24.7 \\
\footnotesize GURA Fuse Cat H+w+C & \checkmark & & \checkmark & & \checkmark & 1339.0 & 3072 & 3072 & 40.1 \\
\footnotesize GURA Fuse Cat H+w+C (time) & \checkmark & \checkmark & & & \checkmark & 1339.0 & 15360 & 3072 & 34.6 \\
\footnotesize GURA Fuse Hubert & \checkmark & & \checkmark & & & 1000.0 & 1280 & 1280 & 18.1 \\
\footnotesize GURA Fuse wav2vec2 & \checkmark & & \checkmark & & & 317.0 & 1024 & 1024 & 8.8 \\
\footnotesize GURA Hubert & \checkmark & & \checkmark & & & 1000.0 & 1280 & 1280 & 17.9 \\
\footnotesize IUT-CSE MLP (audio) & & \checkmark & \checkmark & \checkmark & \checkmark & 0.2 & 1584 & 8 & 2.9 \\
\footnotesize IUT-CSE MLP (keyword) & & \checkmark & \checkmark & & & 0.4 & 1024 & 64 & 3.0 \\
\footnotesize Logitech AI SERAB BYOL-S & & \checkmark & \checkmark & & & 5.3 & 2048 & 2048 & 4.8 \\
\footnotesize RedRice EfficientNet-B2 & & \checkmark & & \checkmark & & 7.7 & 1408 & 1408 & 3.4 \\
\footnotesize Sony UDONS ViT & & \checkmark & \checkmark & & & 11.1 & 768 & 768 & 3.5 \\
\footnotesize Soundsensing YAMNet & & \checkmark & & \checkmark & & 3.8 & 1024 & 1024 & 15.7 \\
\footnotesize Stellenbosch LSL DBERT & \checkmark & & \checkmark & & & 316.8 & 2048 & 2048 & 6.5 \\

\bottomrule
\end{tabular}
  }

\end{table}

\begin{figure}[hbtp]
\floatconts
    {fig:task-correlation}%
    {\caption{Task versus task correlation scores, based upon normalized scores. Only the highest and lowest correlations are displayed. Cells are sorted to minimize the traveling salesperson distance, mapping correlations [-1, +1] to distances [+2, 0].}}%
    {\includegraphics[width=\textwidth]{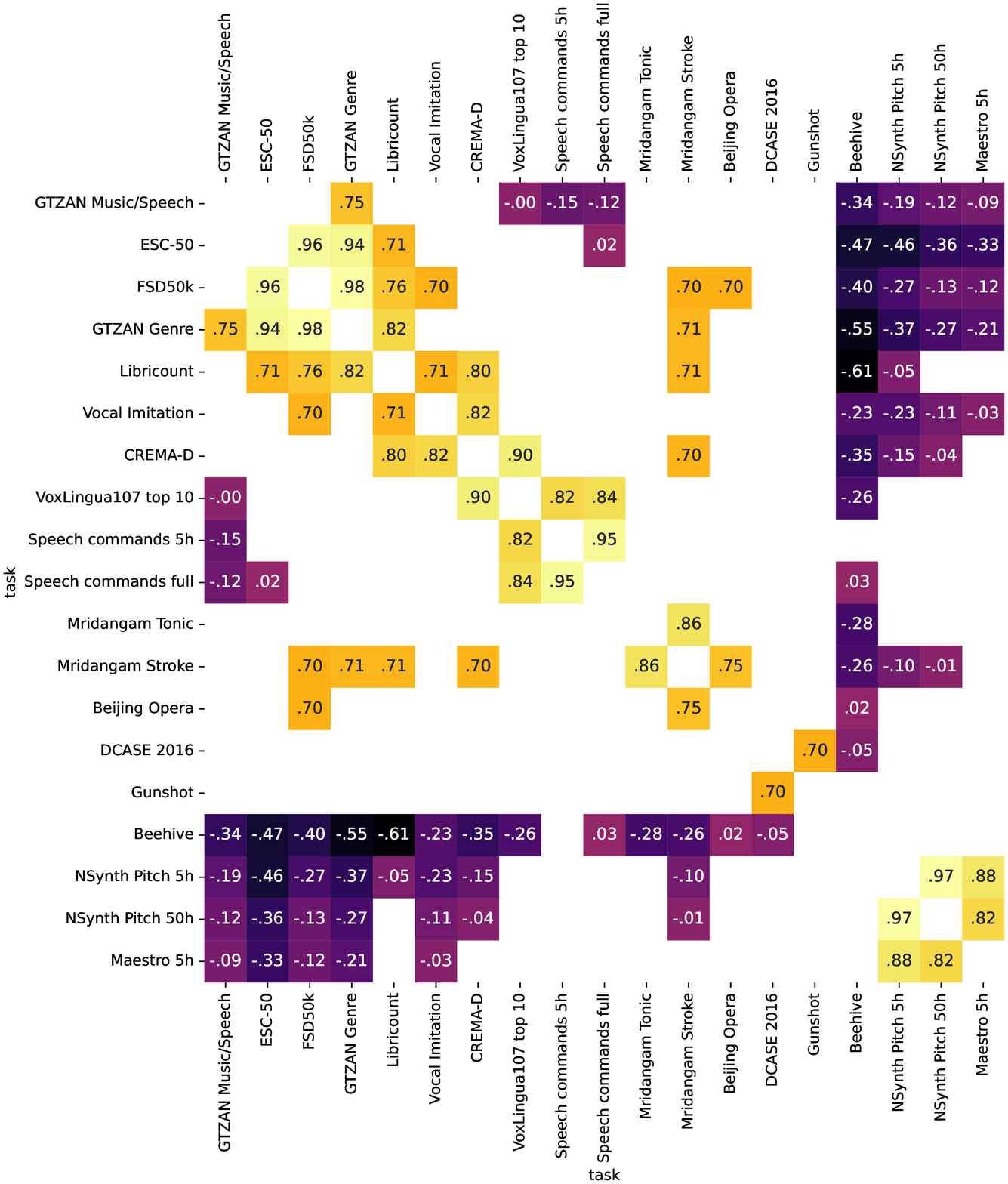}}
\end{figure}

\begin{figure}[hbtp]
\floatconts
    {fig:model-correlation}%
    {\caption{Model versus model correlation scores, based upon normalized scores. Only the highest and lowest correlations are displayed. Cells are sorted to minimize the traveling salesperson distance, mapping correlations [-1, +1] to distances [+2, 0].}}%
    {\includegraphics[width=\textwidth]{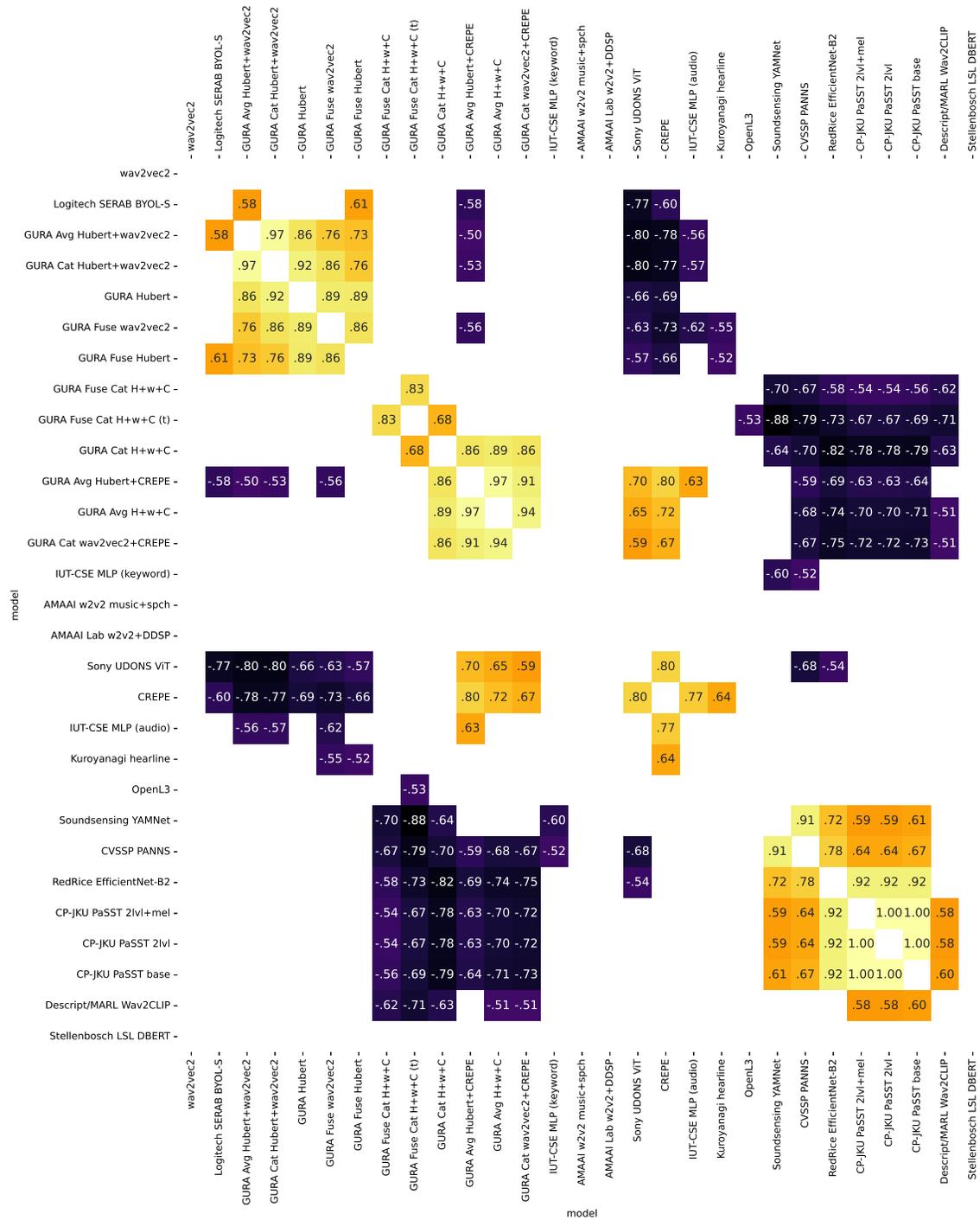}}
\end{figure}

\section{Downstream training details}\label{apd:downstream}

For each task, using a given model's frozen embeddings as input features, we train a downstream MLP classifier. For scene-based multiclass tasks, the final layer is a softmax with cross-entropy loss. For scene-based multilabel tasks and multilabel frame reductions of timestamp tasks, the final layer is a sigmoid with cross-entropy loss.

We monitor the score (not loss) on the validation set. For timestamp tasks, computing the validation score involves a full CPU-based \texttt{sed\_eval} \citep{sedeval} run with median filter of 250ms and minimum event duration 125\,ms and 250\,ms. (Both event durations are tried at each validation step and the best hyperparameter is retained for that validation step.) We train for a maximum of 500 epochs, checking the validation score every 3 epochs, early stopping if no improvement is seen after 20 validation steps. For DCASE 2015 Task 2, we check the validation score every 10 epochs.

The validation score is used for early-stopping, as well as for model selection. The same RNG seed is used for every model-task downstream training, ensuring that grid points and weight initialization is identical. Model selection is performed over 8 deterministic random grid points out of 16 possible grid points. Hyperparameters are shown in \tableref{tab:hyperparameters}.
This grid was chosen after using a much larger hyperparameter grid with the three baseline models on the open tasks. In these preliminary hyperparameter grid pruning experiments, the grid was progressively refined by discarding hyperparemeter choices that were not predictive of relatively high model performance, similarly to how \citet{Kelz2016-vy} use tree ensemble learning to prune their hyperparameter grid.

\begin{table}[htbp]
\floatconts
  {tab:hyperparameters}%
  {\caption{Hyperparameters used for training.}}%
  {
\begin{tabular}{rl}
  \toprule
Hidden layers & [1, 2] \\
Hidden dimensions & 1024 \\
Dropout & 0.1 \\
Learning rate & [3.2e-3, 1e-3, 3.2e-4, 1e-4] \\
Batch size & 1024 \\
Hidden norm & Batch Norm \\
Initialization & [Xavier Uniform, Xavier Normal] \citep{pmlr-v9-glorot10a} \\
Optimizer & Adam \\
  \bottomrule
\end{tabular}
  }
\end{table}

\end{document}